# Evidence for current flow anomalies in the irradiated 2D electron system at small magnetic fields

*R.L. Willett, L.N. Pfeiffer, K.W. West*
*Bell Laboratories, Lucent Technologies*

__*Abstract*__

We report experimental results of low temperature magnetotransport in high mobility 2D electron systems exposed to radiation up to 20GHz frequency using a simple dipole configuration. Magnetoresistance oscillations are observed as in previously reported higher frequency radiation on 2D systems, however minima here can be seen to extend to negative biases, and zeroes previously reported are not observed persistently around the full sample perimeters. In addition, under radiation, voltages are observed from internal to external contacts in the absence of applied driving currents not due to simple rectification. These findings are consistent with micro- and macroscopic theoretical pictures of radiation induced transport and current instabilities due to local negative resistivities. However, the temperature dependent development of minima is shown to be severely power dependent, and outside of present theoretical understanding.

Simple irradiation in the GHz range on 2D electron systems has been shown to have remarkable consequences on the transport at low magnetic fields. It was found by Zudov, et. al. [1] that radiation from 30 to 120GHz imposed on a high quality heterostructure resulted in a series of oscillations periodic in $\omega/\omega_c$ with $\omega$ the radiation frequency and $\omega_c$ the cyclotron frequency, using bare GaAs electron mass. Subsequently it was observed by Mani, et al [2] and Zudov, et al [3] that in high mobility samples the minima can form apparent zeroes, with the temperature dependence activated. The series of minima, formed at $\omega/\omega_c = j + \alpha$, $j = 1,2,3\ldots$, $\alpha = ½$ [1] or ¼ [2], and showed for the primary minima (j = 1, 2) activation energies substantially larger than the incident radiation.

Multiple theoretical addresses of these results have been made [4-7], with one microscopic picture [4] developed that describes radiation induced, disorder facilitated magnetotransport oscillations. In this model radiation raises an electron over multiple Landau levels, with disorder scattering providing a preferential boost or reduction to the conductivity dependent upon the relative position of the new high energy level to the adjacent Landau levels. Numerical calculations from the diagrammatic expansions show an oscillatory magnetotransport with the appropriate period of $\omega/\omega_c$. Significantly, these calculations showed that under high radiation power, the conductivity minima could be seen to become negative. Further work by Andreev, et al. [7], described the macroscopic consequences of this negative conductivity. It was shown that current instabilities will result, with current inhomogeneities or circulations possible: current vortices may result under sufficient radiation power.

We report in our experimental work here prominent magneto-oscillations at low frequencies (<20GHz) and in ultra-high mobility samples that demonstrate the essential physics of the previously observed effects. In our studies, however, we study in detail the oscillatory resistance and observe minima consistent with the previously reported

"zeroes", but also note minima in which the voltage drop is distinctly negative. We have mapped the resistivity properties around sample perimeters and find that all samples show variation in minima, with no samples showing all zeroes around the perimeter. We then tested voltage drops from internal contacts to external contacts with radiation but without applied driving current: substantial voltages were observed, consistent with large currents induced by the radiation yet not consistent with a simple rectification of the radiation. These findings are in sum supportive of both the microsopic and macroscopic pictures developed to describe radiation induced magnetoresistance oscillations. However, an important property of the radiation-induced minima is the temperature dependence: the previous experimental work [2,3] showed activated transport consistent with an energy gap in the transport spectrum. We examined the transport minima and found a complicated temperature dependence: hysteresis in the minima are possible, with temperature development not necessarily activated over a large temperature range. In addition, we find all temperature dependence is severely dependent upon the incident radiation power.

The samples used in these experiments all have mobilities in excess of $15 \times 10^6 cm^2$/V-sec., with densities near $2 \times 10^{11} cm^{-2}$ and were taken from 3 different MBE produced GaAs/AlGaAs wafers. Contacting to the 2D electron system is accomplished with diffused In or Ni/Au/Ge contacts. The Ni/Au/Ge contacts have a particular layering scheme that allows highly efficient contacting in that a large percentage of the area covered by the contact diffuses to the 2D layer. Unlike previous work [1-3], radiation is not applied to the sample by rigid wave-guide but rather by a linear dipole antenna hung over the sample. The dipole is greater than 7mm in extent, and the samples range from 400μm to 5mm in largest dimension. Frequencies applied can range continuously from 2 to 20GHz. Measurements were made in a $He^3$ refrigerator providing temperatures down to 280mK with radiation.

The lower frequency range accessed here examines fundamentally the same physics as the higher frequency range (30-120GHz), as shown by the data in Figure 1. Using indium contacts the oscillations are observable down to near 3GHz at this temperature and the apparent zeroes can be observed at near 8GHz. Such a minimum demonstrating this saturation near zero but at 20GHz was examined for its temperature dependence near full incident power: see Figure 1b. It demonstrates an activated temperature dependence with activation energy E, ($\rho \sim e^{-(E/kT)}$) of 6K. This scales well with the results of Zudov et al [3] where with incident radiation of 54GHz, an activation energy of 18K was derived. One novelty at the lower frequencies is an apparent minimum in resistivity at roughly twice the magnetic field value of the principal (j=1) minimum. Beyond this difference, these results indicate that the process of magneto-oscillation due to incident radiation appears continuously over a large frequency range.

One of the findings of this work is that negative bias can be observed across longitudinal resistivity contacts as shown in Figure 2. The measurement configuration here is constant current driven at low frequency (3 to 13Hz) through a van der Paaw configuration of contacts with voltage tapped along one side of the sample. The magneto-transport of Fig. 2a shows distinct negative bias for the minimum at j=1 as the incident radiation power and frequency are increased. In another sample (Fig. 2b), negative bias is again observed but as incident radiation power is increased structure in the center of the minimum appears and the minimum becomes positive. These negative bias minima have been observed in samples from all three wafers examined. The structure of the minima change with thermal cycling. Better contacting of the 2DES allows observation of the negative bias, as our Ni/Au/Ge contacts with uniform metal diffusion to the conducting layer tend to demonstrate the negative bias with higher prevalence. These results of negative bias can be understood in terms of a current path counter to the net current flow occurring locally near the contacts tested. This picture would suggest that not all contacts would necessarily show negative bias, and this is as we have observed – only a limited number of contact pairs demonstrate this type of negative bias on any given sample.

To further understand possible complicated current flow patterns under illumination we have mapped the voltage differences around the periphery of several high mobility samples. Figure 3 shows the results of such a measurement: similar minima *do not* occur around the entire periphery of the sample. In the sample of Figure 3 minima along one side of the sample show the "zeroes" while the opposite side shows a positive bias where the "zeroes" are anticipated. When the B-field direction is reversed the minima on the two sides of the sample are reversed in properties – "zeroes" now occur where the positive bias was previously observed and the vise-versa. This reversal of properties is symmetric about the line determined by the radiation dipole direction. *All* samples examined to date, including the 400μm x 400μm square samples, have demonstrated this lack of zeroes completely around the sample perimeter and the reversal of zero positions with change in magnetic field direction. This type of voltage mapping shows that a homogeneous resistance is not present throughout the samples upon illumination and the reversal of the minima with B-field reversal shows that improper contacting is not the source of the non-"zero" minimum.

An important property to determine is whether significant currents are generated within the samples upon radiation but in the absence of driving currents. To this end we produced a series of samples with contacts internal to the periphery and also standard peripheral contacts. The voltage difference between this internal and an external contact can be measured; with no external current input, if radiation generates a current within the sample that has a net value between those contacts, that Hall voltage generated may be measurable. The results of such measurements are shown in Figure 4: in these the voltage between the labeled contacts is measured with differential input to a preamplifier (PAR 113) over a series of frequency windows, from d.c. to several Hz. A substantial voltage is indeed generated, with features reflecting properties observed in the standard longitudinal resistivity measurement (Fig. 4a), and representing a large induced current. The voltage generated at the j=1 feature corresponds to a current of roughly 5μA flowing between the contacts. At low frequencies (d.c. to 3Hz) a significant rectification signal is present, but as the frequency detection range is raised (0.1 to 3Hz) this rectification signal becomes weaker, particularly at the j=1 feature. This rectification is minimal between peripheral contacts and distinctly different in form from the internal to external contact voltage (Fig. 4c). While all samples examined have shown induced voltages from center to external contacts under radiation, a particularly surprising result is shown in Figure 4d: for this sample, the feature at j=1 is either a minimum or maximum dependent upon the sweep direction of the B-field. This suggests a marked instability in the current paths due to the radiation, with these results in sum showing that large low frequency currents are induced by radiation.

The findings of negative bias, inconsistent oscillatory minima around the sample perimeters, and large induced voltages from center to external contacts under radiation all suggest that substantial current instabilities and anomalous current paths are produced within the 2D gas at low B-field. Such current pathologies are generally consistent with the theoretical macroscopic picture [7] and by inference the microscopic picture [4] developed describing radiation induced photoconductivity and the current instabilities that should result from negative resistivities [4,7]. Further development of these theoretical pictures is needed to account for these specific experimental findings. While we can generally deduce that substantial current variations occur across the sample, it is not possible to attribute these currents to a homogenous response of the 2DES to the radiation, an intrinsic but inhomogeneous response to the radiation, or an extrinsic input that does not allow homogenous radiation of the sample, such as interference from the sample leads.

An open issue within the standing theoretical models is the temperature dependence of the minima. We examined the temperature dependence of the j=1 minimum at 20GHz but for a range of incident radiation powers. A first observation was a B-field sweep hysteresis effecting the minimum, with data taken in both sweep directions for completeness (Fig. 5a). At minimum attenuation of the incident radiation the temperature dependence is activated over two orders of magnitude in resistance. When less power is incident this activated property is present only for one B-field sweep direction, with the low temperature resistivity showing marked deviation. For lower incident powers the activation energies are clearly of lower value, indicating a strong power dependence to the minimum formation. The activation energies are plotted as a function of incident power (Fig. 5c) showing a linear relationship. This power dependence and the presence of hysteresis indicate that the activated transport may not reflect a microscopic property of the system, but rather a process related to macroscopic features in the transport, such as current path formation or switching.

In conclusion, “zeroes” may not be an appropriate description of resistance in radiation exposed samples as these results are not consistent over entire samples, suggesting inhomogeneous current paths. The current instabilities picture is supported by our results, with local reverse current possibly responsible for the negative voltage measured at some peripheral contact pairs. Other tests for current instabilities, such as large internal to external contact potentials, are consistent with this picture. The oscillation minima demonstrate incident power temperature dependence, which further complicates the description of the minima as zeroes. It remains an open question as to what determines this temperature dependence.

We gratefully acknowledge discussions with N. Read, A. Durst, and A. Andreev.

## Figure captions:

Figure 1. Magnetoresistance oscillations induced in high mobility sample A by radiation of frequencies down to 3GHz. Temperature 280mK. The apparent zeroes are observed at frequencies down to less than 9GHz. The inset shows activated temperature dependence for j=1 minimum over more than two orders of magnitude resistance at 20GHz.

Figure 2. Top traces. Sample B showing principal minimum extending to negative bias as frequency and power are increased. Lower traces. Sample C showing negative bias at j=1 minimum where with increasing power the minimum splits and grows to positive bias. Temperature 280mK.

Figure 3. Mapping of voltage drops around sample D perimeter according to the schematic. For reversed magnetic field direction the apparent zeroes will appear in the contact set on the opposite side of the sample for the radiation dipole direction along the driven current direction.

Figure 4. Samples exposed to 20GHz radiation at 280mK. Figure 4a compares standard longitudinal resistance with the voltage induced across internal to external contacts with no current applied externally to sample E measuring voltage over frequency range d.c. to 3Hz. Figure 4b shows sample F internal to external contact voltage measured over several frequency ranges: the radiation induced voltage occurs predominantly at frequency less than 1 Hz. Figure 4c: voltages across different contact configurations with no drive current in sample G, all d.c to 3Hz. Figure 4d: internal to external contact voltage for magnetic field sweeps up and down in sample H showing distinct j=1induced voltages, again d.c. to 3Hz.

Figure 5. Temperature dependence of principal minimum in sample I which shows hysteresis: values in the respective magnetic field sweep directions are shown in the lower plot for different applied radiation powers. The lines through the data are guides for the eye. The activation energies derived from the up sweeps show a linear increase in value for increasing power (inset).

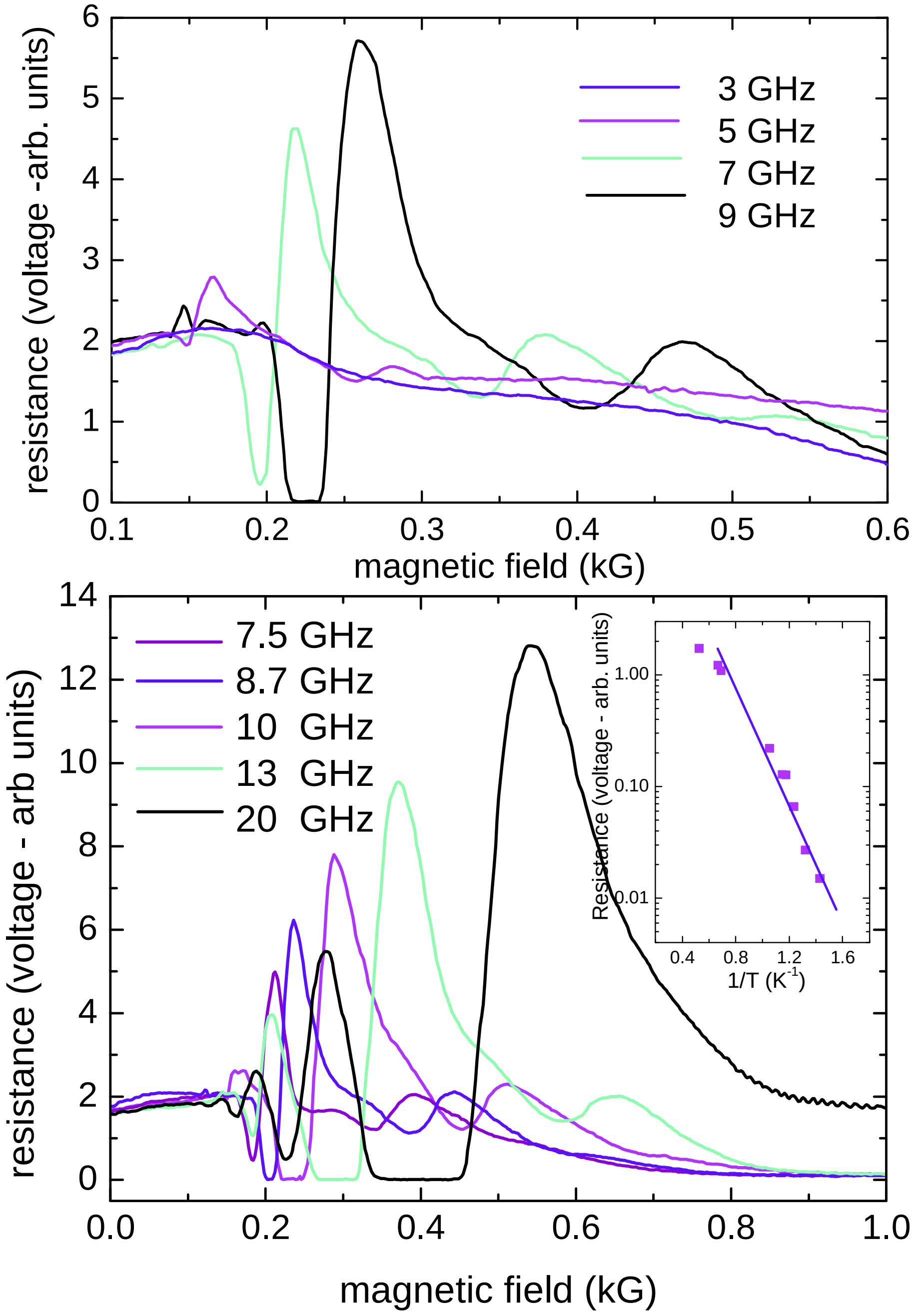

Figure 1

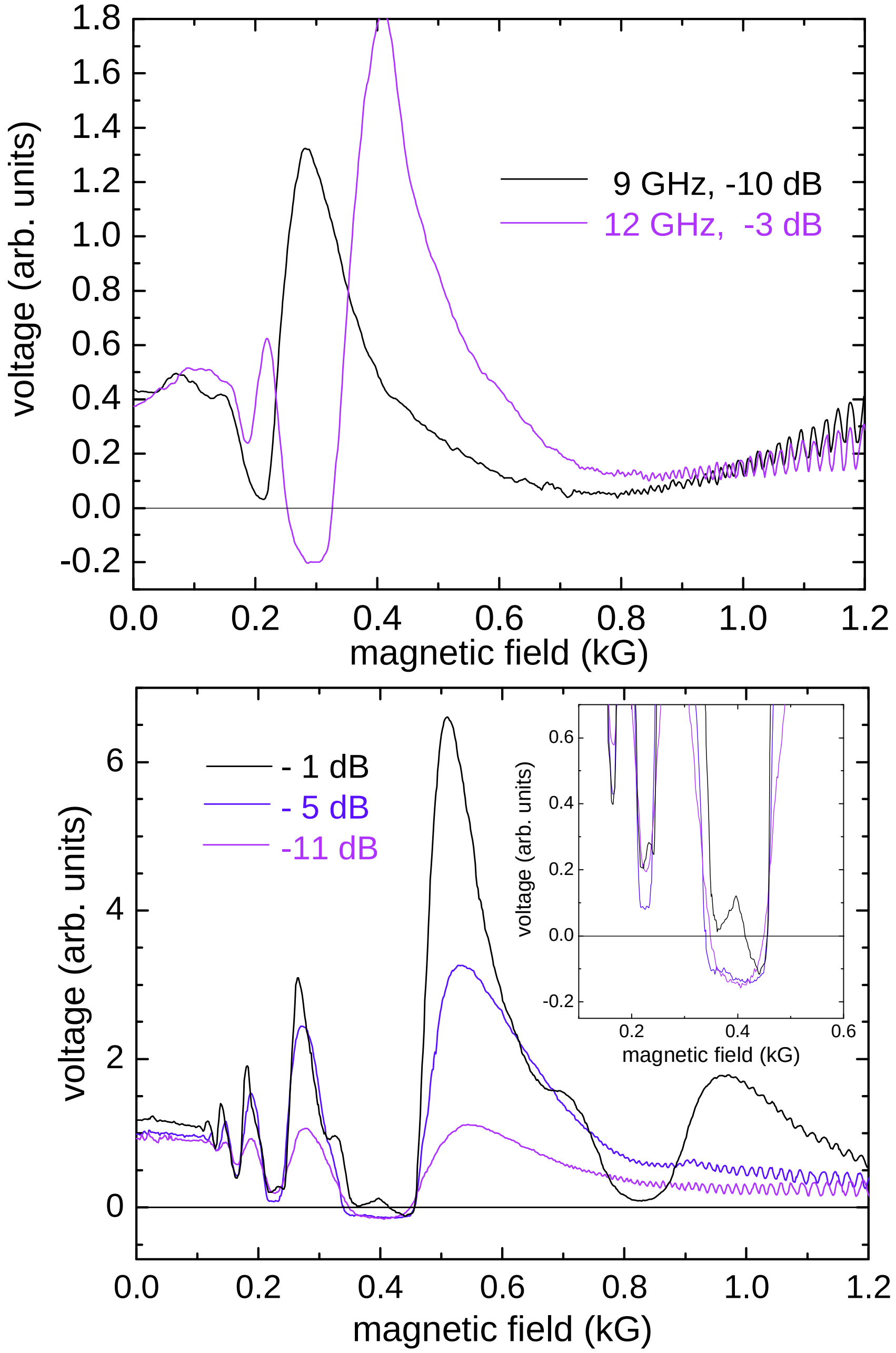

voltage (arb. units)
magnetic field (kG)
9 GHz, -10 dB
12 GHz, -3 dB
- 1 dB
- 5 dB
-11 dB

Figure 2

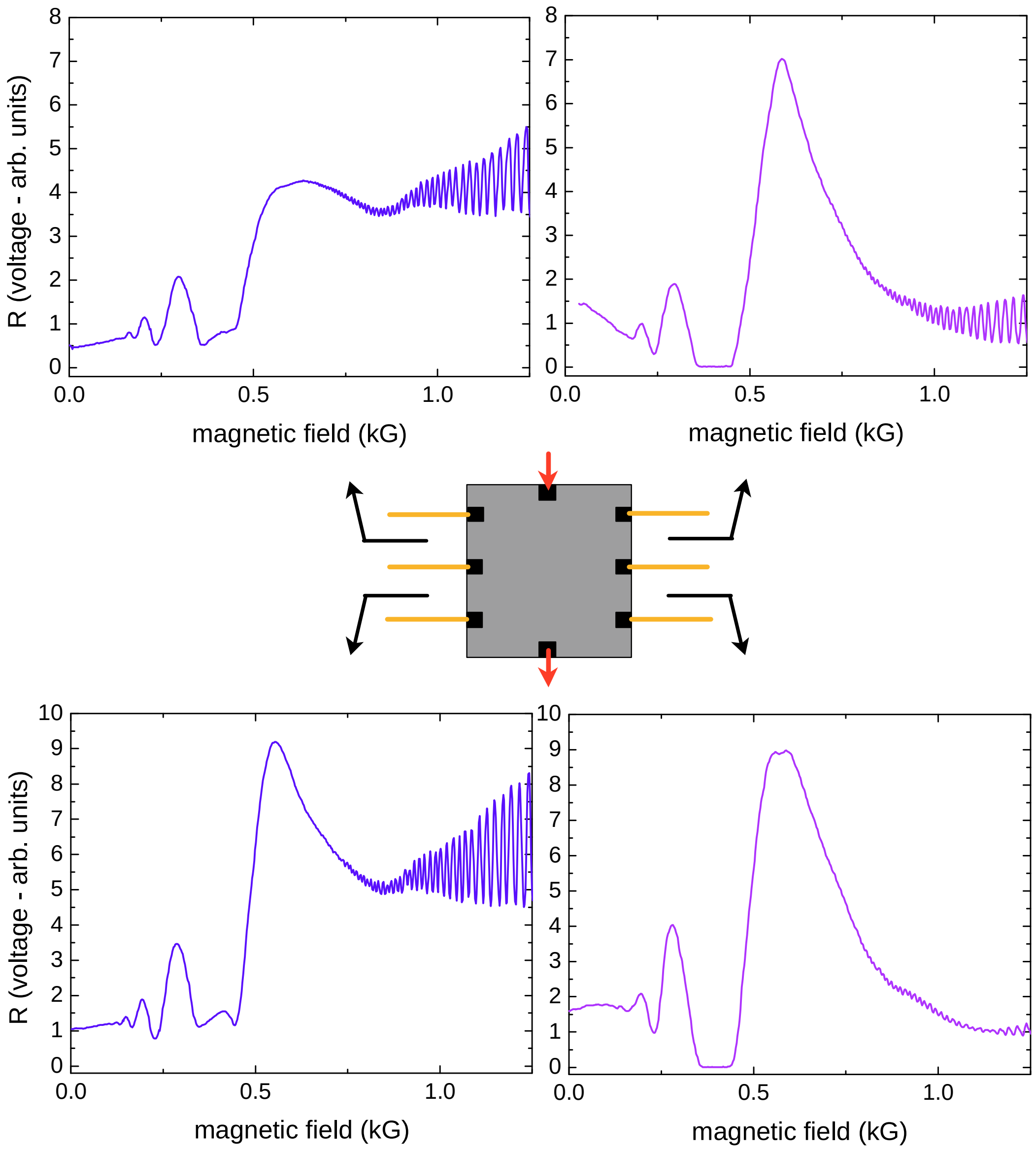

Figure 3

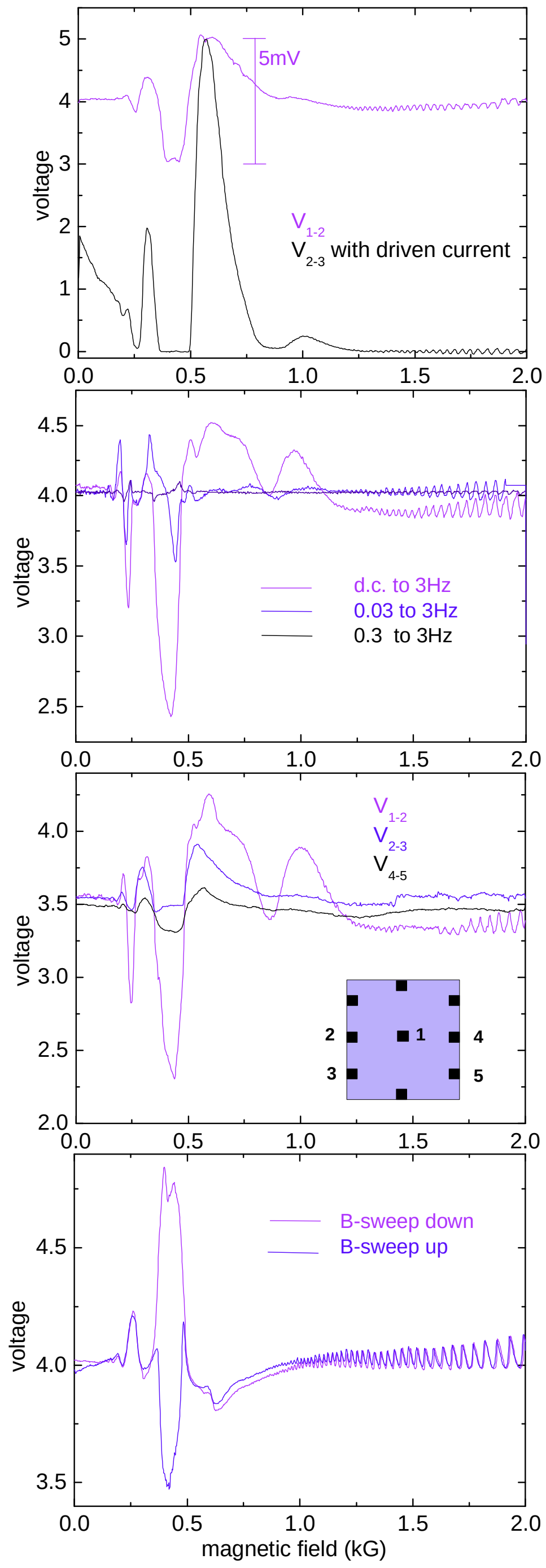

Figure 4

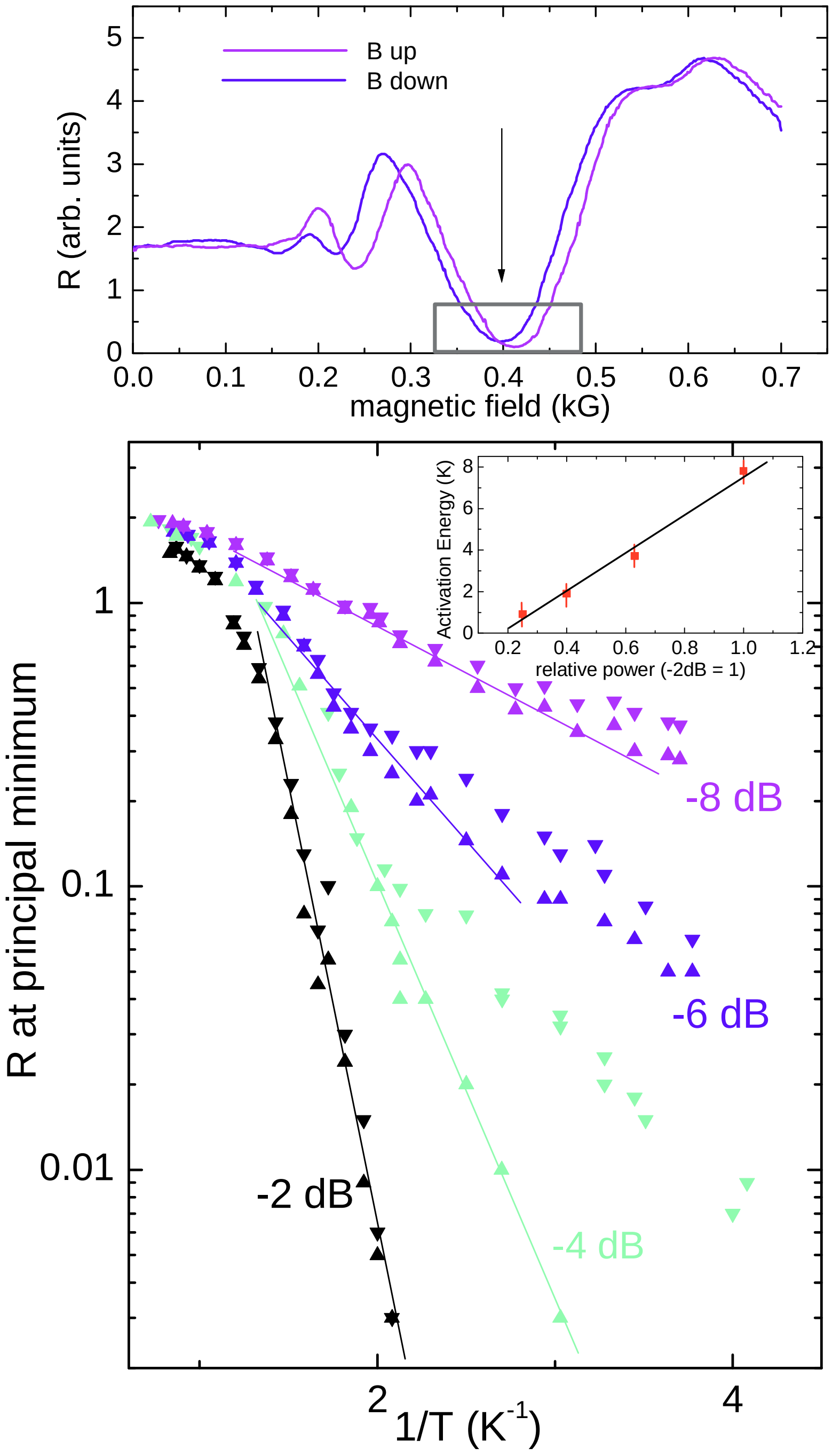

Figure 5